\documentclass[aps,prc,letterpaper,reprint,twocolumn,superscriptaddress,showpacs,amsmath,amssymb]{revtex4-2}

\usepackage{graphicx}
\usepackage{amsmath}
\usepackage{braket}
\usepackage{url}
\usepackage{multirow}
\usepackage{amssymb} 
\usepackage{booktabs}
\usepackage{rotating}
\usepackage{bm}
\usepackage{bbm}
\usepackage{longtable}
\usepackage{subfigure}
\usepackage{tikz}
\usetikzlibrary{matrix}

\usepackage{amssymb}

\usepackage{braket}
\usepackage{amsmath}
\usepackage{graphicx}
\usepackage{comment}

\usepackage[colorlinks,
linkcolor=blue,
anchorcolor=red,
urlcolor=red,
citecolor=red]{hyperref}
\usepackage{cancel}

\begin{document}


\title{Gamow shell model calculations for the Thomas-Ehrman shift in new isotopes $^{21}$Al}

\author{K. H. Li}
\thanks{These authors contributed equally to this work.}
\affiliation{CAS Key Laboratory of High Precision Nuclear Spectroscopy, Institute of Modern Physics, Chinese Academy of Sciences, Lanzhou 730000, China}
\affiliation{College of Physics, Center for Theoretical Physics, Henan Normal University, Xinxiang 453007, China}

\author{N. Chen}
\thanks{These authors contributed equally to this work.}
\affiliation{CAS Key Laboratory of High Precision Nuclear Spectroscopy, Institute of Modern Physics, Chinese Academy of Sciences, Lanzhou 730000, China}
\affiliation{School of Nuclear Science and Technology, University of Chinese Academy of Sciences, Beijing 100049, China}

\author{J. G. Li}
\email{jianguo\_li@impcas.ac.cn} 
\affiliation{CAS Key Laboratory of High Precision Nuclear Spectroscopy, Institute of Modern Physics, Chinese Academy of Sciences, Lanzhou 730000, China}
\affiliation{School of Nuclear Science and Technology, University of Chinese Academy of Sciences, Beijing 100049, China}

\author{H. H. Li}
\affiliation{CAS Key Laboratory of High Precision Nuclear Spectroscopy, Institute of Modern Physics, Chinese Academy of Sciences, Lanzhou 730000, China}
\affiliation{School of Nuclear Science and Technology, University of Chinese Academy of Sciences, Beijing 100049, China}

\author{M. R. Xie}
\affiliation{CAS Key Laboratory of High Precision Nuclear Spectroscopy, Institute of Modern Physics, Chinese Academy of Sciences, Lanzhou 730000, China}

\author{C. W. Ma}
\affiliation{College of Physics, Henan Normal University, Xinxiang 453007, China}
\affiliation{Institute of Nuclear Science and Technology, Henan Academy of Sciences, Zhengzhou 450015, China}

\author{W. Zuo}
\affiliation{CAS Key Laboratory of High Precision Nuclear Spectroscopy, Institute of Modern Physics, Chinese Academy of Sciences, Lanzhou 730000, China}
\affiliation{School of Nuclear Science and Technology, University of Chinese Academy of Sciences, Beijing 100049, China}

\date{\today}

\begin{abstract}

Proton-rich nuclei beyond the proton drip line exhibit unique phenomena, such as the Thomas-Ehrman shift (TES), providing valuable insights into nuclear stability and isospin symmetry breaking.  The discovery of the lightest new isotope, $^{21}$Al, situated beyond the proton drip line, was recently reported in the experiment. In this study, we employ the Gamow shell model (GSM) to explore the TES in mirror pairs $^{21}$Al/$^{21}$O, focusing on how this phenomenon affects the energy levels of these nuclei.
Our calculations describe the ground state energies and reveal significant TES with large mirror energy differences in the excited mirror $1/2^+$ states in $^{21}$Al/$^{21}$O. The large mirror energy difference is primarily due to the significant occupation of weakly bound or unbound $s_{1/2}$ orbitals, resulting in the extended radial density distributions, and variations in Coulomb energy and nuclear interaction contributions between the mirror states.
Additionally,  the low-lying states of $^{21}$Al are also calculated with the GSM in coupled-channel (GSM-CC) representation, Furthermore, we also predict the cross-section of $^{20}$Mg(p, p) scattering, which serves as another candidate approach to study the unbound structure of $^{21}$Al in the experiment,  offering a theoretical framework for studying the structure and reaction dynamics of $^{21}$Al in future experiments.

\end{abstract}

\maketitle

\section{Introduction}  

The exploration of unbound proton-rich nuclei beyond the proton drip line, which primarily decay via proton(s) emission, has emerged as a pivotal area of research in nuclear physics, offering profound insights into the limits of nuclear stability and the underlying mechanisms governing nuclear interactions. 
The study of these nuclei has gained significant momentum with the experimental discovery of various proton-emission modes, such as $1p$ decay of $^{15}$F~\cite{PhysRevC.69.031302}, $2p$ decay of $^6$Be~\cite{PhysRevLett.109.202502}, $^{19}$Mg~\cite{PhysRevLett.99.182501} and $^{16}$Ne~\cite{PhysRevLett.113.232501}, $3p$ decay of $^7$B~\cite{PhysRevC.84.014320} and $^{13}$F~\cite{PhysRevLett.126.132501}, $4p$ decay of $^{8}$C~\cite{PhysRevLett.32.1207,PhysRevC.82.041304} and $^{18}$Mg~\cite{PhysRevLett.127.262502,PhysRevC.103.044319}, and $5p$ decays of  $^9$N~\cite{PhysRevLett.131.172501}.
The study of these nuclei is crucial for enhancing our understanding of nuclear structure under extreme proton-rich conditions, which has implications for fundamental nuclear theories.


Unbound proton-rich nuclei, toward the limits of nuclear structure, serve as examples of open quantum systems.
These nuclei exhibit anomalous phenomena influenced by their coupling to the continuum of scattering states. One of the most significant is the Thomas-Ehrman shift (TES)~\cite{PhysRev.88.1109,PhysRev.81.412}. It is initially proposed to explain the unexpectedly mirror energy difference (MED) of the $1/2^{+}$ state in $^{13}$C compared to its mirror nucleus $^{13}$N~\cite{PhysRev.88.1109,PhysRev.81.412}.
TES manifests as shifts in energy levels and modifications in decay properties~\cite{PhysRevC.108.064316,OGAWA1999157,Hoff2020,PhysRevC.91.024325}, thereby providing critical tests for nuclear models. Understanding TES in such systems is essential for accurately describing the behavior of proton-rich nuclei and for predicting the existence and properties of yet-undiscovered isotopes~\cite{PhysRevLett.127.262502}.
Numerous states exhibiting large TES with their mirror partners, $^{18}$Ne~\cite{PhysRevLett.102.152502,ZHANG2022136958}, $^{19}$Na~\cite{PhysRevC.67.014308}, and $^{22}$Al~\cite{PhysRevLett.125.192503,Sun_2024} serve as notable examples.
Theoretical description of the properties of nuclei beyond drip lines is one of the main challenges. Furthermore, Coulomb and continuum effects are competitive in the unbound light proton-rich nuclei, which both affect spectroscopic properties such as the ground states inversion that occurs in $^{16}$F/$^{16}$N~\cite{PhysRevC.90.014307,PhysRevC.106.L011301} mirror nuclei.

For the TES, two distinct theoretical interpretations have been proposed. The first explanation attributes the phenomenon to valence proton occupying weakly bound or unbound $s$ single-particle state. In such cases, the single-particle wavefunctions of the $s$  orbital exhibit significantly larger spatial extension compared to those of neutron single-particle wavefunctions. 
For example, it was shown that the shift of the TES in $1/2^{+}$ state of $^{13}$N/$^{13}$C is mainly caused by the different nature of the radial parts of the $s$ and $p$ wave single-particle functions, which principally affect the Coulomb matrix elements~\cite{PhysRevC.63.017301} (the detail of the theory in Ref.~\cite{AUERBACH1983273}). 
The second interpretation operates within the configuration mixing framework, where the continuum coupling is explicitly included. In this framework, the many-body states are described by mixing Slater determinants. The state, located nearby the particle emission threshold, exhibits significant coupling with the continuum states, resulting in more spatially extended wave function compared to the bound states, particularly for valence proton mainly in $s$ waves. Consequently, mirror nuclear states develop different asymptotic behaviors in their wavefunction tails, leading to the energy shifts of the states in proton-rich nuclei compared to those in neutron-rich nuclei.

A new isotope beyond the proton drip line, $^{21}$Al has been observed in the experiment~\cite{PhysRevC.110.L031301}.
By measuring the angular correlations of decay products from the unbound nucleus $^{21}$Al, which spontaneously emits 1$p$, researchers accurately determined the 1$p$ decay energies for its ground and low-lying excited states, as well as the proton separation energy relative to $^{20}$Mg. Additionally, $^{20}$Mg(p, p) scattering experiments offer an alternative method for probing the structure of $^{21}$Al, which could provide complementary information on its excitation spectrum and reaction dynamics. 

Since $^{21}$Al is unbound, it is difficult to study by traditional shell models, necessitating a more sophisticated theoretical framework. The Gamow Shell Model (GSM) ~\cite{PhysRevLett.89.042502,PhysRevC.84.051304,PhysRevC.88.044318,physics3040062,PhysRevC.106.L011301,Michel_2009,PhysRevC.67.054311,PhysRevC.96.024308} is an extension of the conventional shell model. The GSM works in the Berggren basis, which integrates both bound state, continuum coupling and resonance, is ideally suited for these applications. 
Indeed, GSM has become a reliable predictive tool for describing the unbound properties of dripline nuclei~\cite{PhysRevC.106.L011301,Michel_2009,PhysRevLett.89.042502,PhysRevC.67.054311,PhysRevC.96.024308,Li2024,PhysRevC.103.034305}, as well as the TES in proton-dripline nuclei~\cite{PhysRevC.102.024309,PhysRevC.108.064316,PhysRevC.106.L011301,PhysRevC.104.024319}. This capability has been demonstrated in studies of $^{16}$Ne and $^{18}$Mg~\cite{PhysRevC.103.044319}, the $A = 16$ mirror pairs $^{16}$F/$^{16}$N and $^{16}$Ne/$^{16}$C~\cite{PhysRevC.108.064316}, and the mirror nuclei of oxygen isotopes~\cite{GSM_MED}.
Furthermore, the GSM with coupled-channel approach (GSM-CC) enables the calculation of $^{20}$Mg(p, p) scattering cross sections using the same Hamiltonian within the GSM framework, providing a unified theoretical framework to describe both nuclear structure and reaction phenomena.

In this paper, we present a theoretical study on $^{21}$Al. This paper is structured as follows. First, the basic features of GSM and GSM-CC methods are briefly introduced. Then, the low-lying states of mirror partners $^{21}$Al/$^{21}$O and TES are investigated by the GSM calculations. Afterward, the low-lying state of $^{21}$Al and the $^{20}$Mg(p, p) scattering cross section are calculated by the GSM-CC. Finally, conclusions related to exotic structure and reactions for the unbound $^{21}$Al are made.

 \section{Theoretical framework} 

The GSM work in a multi-configuration interaction framework within the the Berggren basis~\cite{PhysRevC.47.768,BERGGREN1968265}, where a core plus valence nucleon(s) picture is adopted~\cite{PhysRevLett.89.042501,PhysRevLett.89.042502,Michel_2009}. The Berggren basis~\cite{PhysRevC.47.768,BERGGREN1968265} includes bound, resonance, complex-energy scattering states, and compose a complete set of states:
\begin{equation}\label{eq1}
\sum_{n}\left|u_{n}\right\rangle\left\langle u_{n}\right|+\int_{{L^{+}}}\left|u(k)\right\rangle\left\langle u(k)\right|dk=\mathbf{\hat{1}},
\end{equation}
where the $\left|u_{n}\right\rangle$ is bound or resonance one-body state and $\left|u(k)\right\rangle$ is scattering state belonging to the $L^+$ contour of complex momenta. The $L^+$ contour must encompass all the resonances present in the discrete sum of Eq.(\ref{eq1}) (the details in Ref.~\cite{Michel_2009}). Such that, this approach allows for a proper treatment of both internucleon correlations and continuum coupling. The GSM Hamiltonian, defined by its complex symmetry, includes numerous many-body scattering states. To diagonalize the GSM Hamiltonian matrix and identify resonances among the many-body continuum states, the Jacobi-Davidson~\cite{MICHEL2020106978,Jacobi-davidson_methods,Michel_2009} and the overlap method~\cite{PhysRevC.67.054311,Michel_2009} are uesed in practical GSM calculations.

The interaction between the core and valence nucleons is modeled via a one-body Woods-Saxon (WS) potential, while the nucleon-nucleon interaction among valence nucleons is described by effective field theory (EFT) forces~\cite{MACHLEIDT20111,RevModPhys.81.1773,RevModPhys.85.197,RevModPhys.92.025004}, in which the three-body counterterm at leading order was ignored in this work. The Coulomb interaction is included for valence protons. Moreover, the effects of three-body interactions are taken into account through the $A$-dependent Hamiltonian, as done in Ref.~\cite{PhysRevC.100.064303}.


In this work, we used $^{16}$O as an inner core, the $s_{1/2}$, $p_{1/2,3/2}$, and $d_{3/2,5/2}$ partial waves are considered by Berggren basis, in which the $L^+$ contour are discretized with 30 points each. The $f_{5/2,7/2}$ are calculated by HO basis with 7 HO states for each partial wave, as the $l$ is larger than $spd$, centrifugal barrier is higher, and contribution to continuum effects is weaker. Moreover, we used the Berggren basis which allows at most two protons or neutrons occupying scattering states to generate natural orbital basis, and calculated physical quantities within natural orbital basis which allows at most three protons or neutrons occupying scattering states. The WS core potential has a diffuseness $d$ = 0.65 fm, a radius $R_0$ = 2.98 fm, and potential depths depend on orbital angular momentum: $V_o$ = 57.50, 58.07, 58.00, and 58.07 MeV, respectively for $l$ = 0, 1, 2, 3. Spin-orbit potential depths read: $V_{so}$ = 6.54 MeV for all partial waves. 
The optimized EFT interaction, which could reproduce the experimental spectra reported in Ref.~\cite{arxiv} is adopted. The $A$ dependence of two-body interaction, as done in Ref.~\cite{PhysRevC.100.064303}, is adopted, where $e = 0.3$ is used in the present work. 


In order to connect the nuclear structure to the reaction process, the  the Gamow shell model in coupled-channel (GSM-CC) representation has been adopted~\cite{Michel_Springer,PhysRevC.106.L011301,PhysRevC.89.034624,PhysRevC.99.044606,PhysRevC.91.034609}, which can explicitly handles the coupling between different reaction channels. The GSM-CC equations are derived by expanding the GSM eigenstates in a complete basis of channel states \begin{equation}\label{eq4}|(c, r)\rangle=\hat{\mathcal{A}}\left\{\left|\Psi_{\mathrm{T}}^{J_{\mathrm{T}}}\right\rangle \otimes|r \ell j\rangle\right\}_{M_{\mathrm{A}}}^{J_{\mathrm{A}}},
\end{equation}
where $|r \ell j\rangle$ represents a nucleon projectile state and \(|\Psi^{J_T}_{M_T}\rangle\) represents the \(A-1\) target state, which is an eigenstate of the GSM Hamiltonian. 
The reaction dynamics are then characterized by the relative motion between the target and projectile nuclei and the channel parameters.


The coupled-channel equations are derived from the Schrödinger equation:
\begin{equation}\label{eq5}\sum_{c} \int_{0}^{+\infty}\left(H_{c c^{\prime}}\left(r, r^{\prime}\right)-E N_{c c^{\prime}}\left(r, r^{\prime}\right)\right) u_{c}\left(r^{\prime}\right) d r^{\prime}=0,
\end{equation}
with 
\begin{equation}\label{eq6}H_{c c^{\prime}}\left(r, r^{\prime}\right)=r r^{\prime}\left\langle(c, r)|\hat{H}|\left(c^{\prime}, r^{\prime}\right)\right\rangle,
\end{equation}
\begin{equation}\label{eq7}
N_{c c^{\prime}}\left(r, r^{\prime}\right)=r r^{\prime}\left\langle(c, r)|\left(c^{\prime}, r^{\prime}\right)\right\rangle,
\end{equation}
where matrix elements for channel coupling and normalization \textcolor{blue}{are} computed via Berggren basis expansion. 
Channel non-orthogonality is addressed by diagonalizing the overlap matrix, transforming the equations into a standard reaction form. The radial wave functions are then solved through integro-differential equations involving local and non-local potentials. This method efficiently describes both bound and resonance states, making it ideal for complex nuclear systems. For a detailed presentation of the GSM-CC method, we refer the reader to Ref.~\cite{Michel_Springer}.

\begin{figure}
    \centering
    \includegraphics[width=0.88\columnwidth]{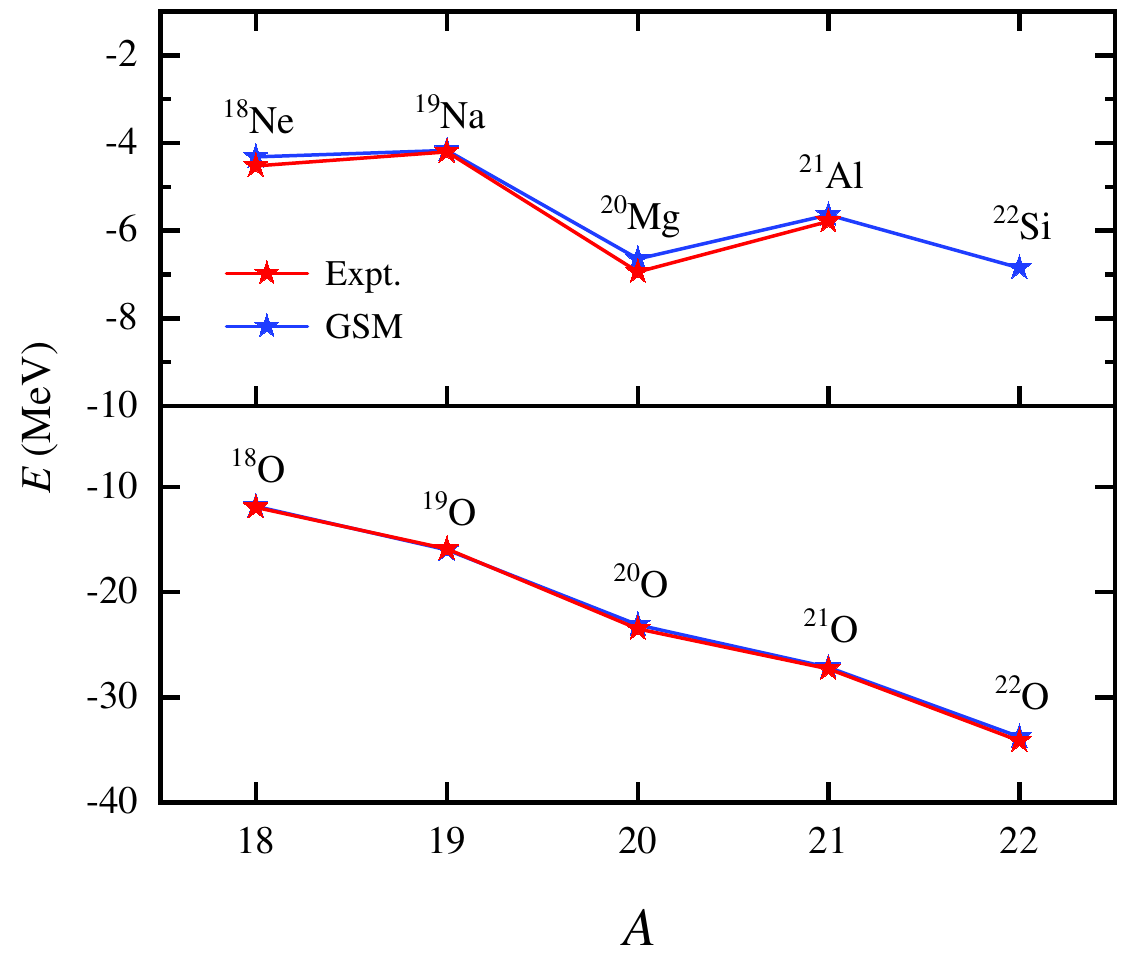}
    \caption{The ground state energy of \textit{A} $= 18-22$ isotones and their mirror partners with $^{16}$O core. Experimental data for $^{21}$Al have been available ~\cite{PhysRevC.110.L031301}, while other data are taken from ~\cite{ensdf}.}
    \label{21Al_22Si_Absolute_Energy}
\end{figure}

\section {Results} 

In Fig~\ref{21Al_22Si_Absolute_Energy}, the ground state energies of $A=18-22$ isotones and their mirror partners with $^{16}$O core were calculated, it was observed that GSM calculations reproduce the ground state energies of nuclei in this region. For $^{22}$Si which has not available experimental mass, the two-proton separation energy ($S_{2p}$) $-108$ (125) keV was given by indirect mass measurements~\cite{XU2017312}, $-120$ keV by three-nucleon forces calculation~\cite{PhysRevLett.110.022502} and $-104$ (201) keV by the compilation of AME2003~\cite{AUDI20033}. But in our calculation, detailed continuum effects were fully taken into account, resulting in a bound ground state for $^{22}$Si with $S_{2p}$ = 172 keV. Moreover, the 1.03 MeV above $^{20}$Mg ground state energy was obtained through GSM for the ground state of $^{21}$Al,  which shows a strong agreement with the experimentally measured value of 1.15 MeV~\cite{PhysRevC.110.L031301}. For other low-lying states, our GSM calculations describe the excitation energies of the mirror partners $^{21}$Al/$^{21}$O in Table~\ref{tab:my_labe 1}, with the maximal deviation from experiment is around 300 keV. 
The excitation energies of ${1/2}^+$ and ${3/2}^+$ in $^{21}$Al predicted by GSM calculations are also close to the \textit{ab initio} calculation with $NN+3N$ forces within $sdf_{7/2}p_{3/2}$ valence space calculation being about 0.5 and 1.4 MeV, respectively~\cite{PhysRevLett.110.022502}. Furthermore, a width of 126 keV is predicted by GSM for the ${1/2}^+$ state of $^{21}$Al.

\begin{table}[!htb]
    \renewcommand{\arraystretch}{1.0}
    \setlength{\tabcolsep}{1.8mm}
    \caption{The calculated excitation energies (in MeV) and MED (all in keV) of the mirror partners $^{21}$Al/$^{21}$O with the GSM, along with available experimental data~\cite{PhysRevC.110.L031301,ensdf}. }
    \label{tab:my_labe 1}
    \begin{tabular}{cccccccccc}
    \toprule

    $J^\pi$ & $E_{\rm exp}$ & $E_{\rm GSM}$ &~ & $E_{\rm exp}$ & $E_{\rm GSM}$ & ~& EXP & GSM\\
    \midrule
    
    & \multicolumn{2}{c}{$^{21}$Al} & ~ & \multicolumn{2}{c}{$^{21}$O} &~& \multicolumn{2}{c}{MED}\\
    \cmidrule{2-3}\cmidrule(r){5-6} \cmidrule(r){8-9}
    $5/2_1^+$ & 0 & 0 &~& 0 & 0  &~& 0 & 0\\
    $1/2_1^+$ & 0.35 & 0.61 & ~ & 1.22 & 1.37 &~& $-870$ & $-760$\\
    $3/2_1^+$ &  & 1.37 & ~ & 2.13 & 1.95 &~&  & $-580$\\
    \bottomrule
    \end{tabular}
\end{table}

\begin{table}[]
    \renewcommand{\arraystretch}{1.2}
    \setlength{\tabcolsep}{4.0mm}
    \caption{The calculated main configurations and their probabilities for the low-lying states in $^{21}$Al/$^{21}$O mirror nuclei with GSM.}
    \label{tab:my_labe 3}
    \begin{tabular}{cccccc}
    \toprule
   \multirow{2}*{$J^\pi$ } &\multirow{2}*{Configurations}&\multicolumn{2}{c}{Probabilities}\\
    \cmidrule{3-4}
    & & $^{21}$Al & $^{21}$O\\
    \midrule
    {${5/2}_1^+$}&$(0d_{5/2})^5$&80\%&82\%\\
    &$(1s_{1/2})^2(0d_{5/2})^3$&6\%&5\%\\
    &$(0d_{5/2})^3(0d_{3/2})^2$&5\%&5\%\\
    {${1/2}_1^+$}&$(1s_{1/2})^1(0d_{5/2})^4$&87\%&87\%\\
    &$(1s_{1/2})^1(0d_{5/2})^2(0d_{3/2})^2$&8\%&8\%\\
    
    {${3/2}_1^+$}&$(1s_{1/2})^1(0d_{5/2})^4$&71\%&72\%\\&$(1s_{1/2})^2(0d_{5/2})^3$&19\%&19\%\\
    \bottomrule
    \end{tabular}
    
\end{table}

\begin{table}[]
    \centering
    \setlength{\tabcolsep}{2.0mm}
    \caption{The calculated average occupations of the $s_{1/2}$, $d_{5/2}$ and $d_{3/2}$ partial waves for the low-lying states of $^{21}$Al/$^{21}$O. }
    \label{tab:my_labe 4}
    \begin{tabular}{ccccccc}
    \hline\hline
    \multirow{2}*{$J^\pi$ } & \multicolumn{3}{c}{$^{21}$Al} & \multicolumn{3}{c}{$^{21}$O} \\
    & $\pi s_{1/2}$ & $\pi d_{5/2}$ & $\pi d_{3/2}$ &$\nu s_{1/2}$ & $\nu d_{5/2}$ & $\nu d_{3/2}$\\
    \hline
    $5/2_1^+$ & 0.21 & 4.64 & 0.15 & 0.16 & 4.68 &0.15\\
    $1/2_1^+$ & 1.03 & 3.76 &  0.20 & 1.00 &3.79 &0.20\\
    $3/2_1^+$ & 1.22 & 3.67 & 0.10 & 1.20 &3.69 & 0.11\\
    \hline\hline
    \end{tabular}
\end{table}

To investigate the TES in these mirror nuclei, we define the MED for a given state $J^{\pi}$: $\mathrm{MED}(T,J^\pi)=E_{x}\left(J^\pi,T_{z_>}\right)-E_{x}\left(J^\pi,T_{z_<}\right)$, in which the $E_{x}$ is excitation energy and $T_{z_>}$($T_{z_<}$) refers to the nucleus of largest(smallest) isospin in mirror nuclei considered in the MED calculations. The MED of GSM calculations and experiments are also given. It is observed that the calculated MED values for ${1/2}^+$ states in $^{21}$Al/$^{21}$O align well with the experimental data.  $^{21}$Al/$^{21}$O highlights significant TES in low-lying states, as their large MED values.

\begin{figure}[htb!]
    \centering
    \includegraphics[width=0.9\columnwidth]{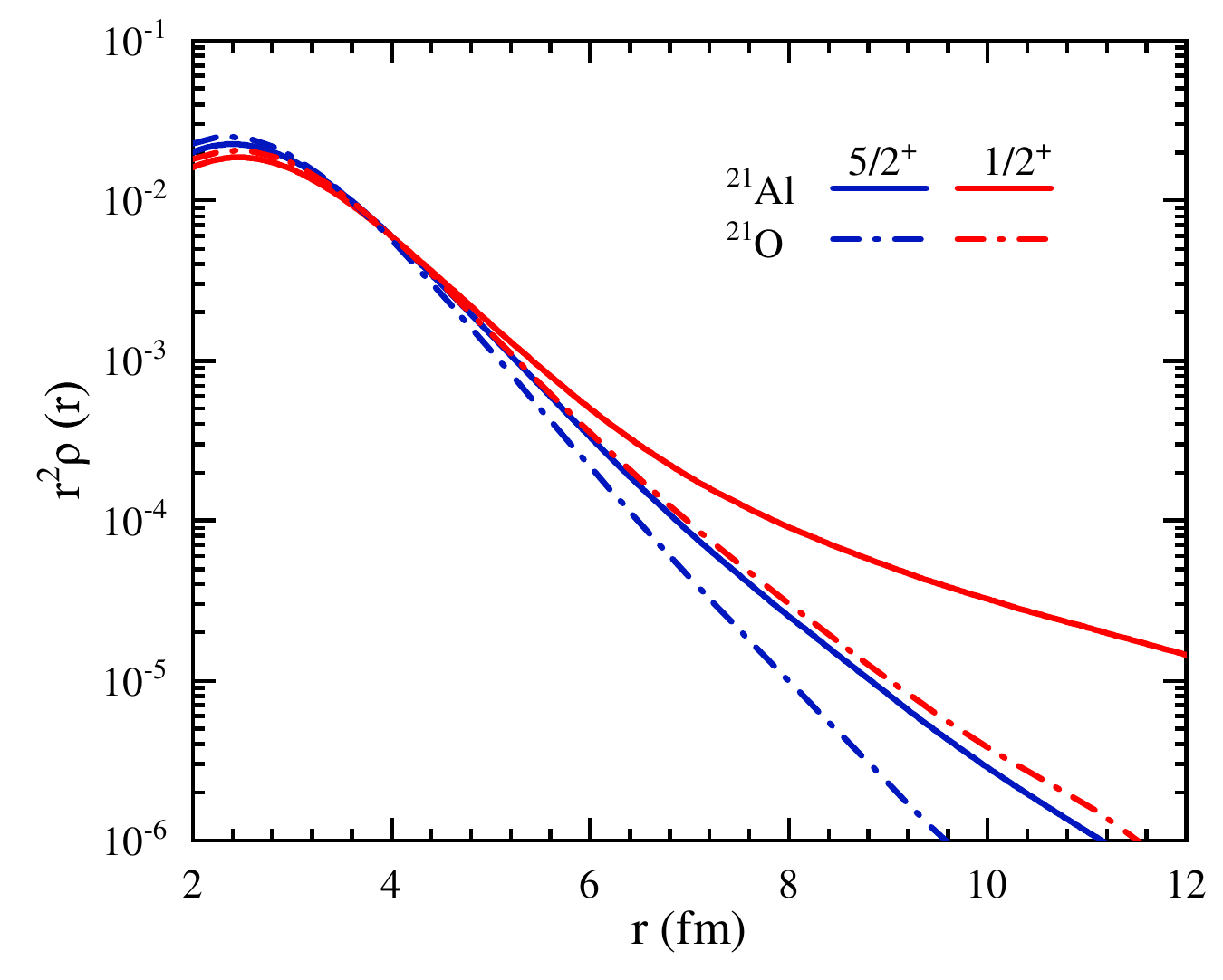}
    \caption{The radial density distributions of the ground and first excited states in $^{21}$Al/$^{21}$O are shown, with the solid line representing $^{21}$Al and the dotted line representing $^{21}$O.}
    \label{21Al_21O_density}
\end{figure}

To explore the significant isospin symmetry breaking~\cite{PhysRevLett.89.142502,PhysRevLett.92.132502,PhysRevLett.97.132501,PhysRevLett.97.152501,PhysRevLett.110.172505,KANEKO2017521} and the corresponding large MEDs, we calculate the configurations, average occupations and radial density distributions of low-lying states by GSM in Tables.~\ref{tab:my_labe 3} and ~\ref{tab:my_labe 4} and Fig.~\ref{21Al_21O_density}.
The many-body configurations and their probabilities for the mirror partners $^{21}$Al/$^{21}$O are almost identical, consistent with the general trend observed for nuclei in the $sd-$shell exhibiting large MED~\cite{10.1088/1674-1137/acf035,PhysRevC.107.014302}. The ground state configurations of both $^{21}$Al and $^{21}$O are predominantly governed by the $(0d_{5/2})^5$, which are also evident in the calculated average occupation, mainly occupied by the $d_{5/2}$ orbital. Although $^{21}$Al is unbound, it is still constrained by a strong centrifugal barrier and Coulomb barrier. This results in the phenomenon as radial density distributions are nearly identical, but drop off rapidly in the asymptotic regions, as shown in Fig.~\ref{21Al_21O_density}. However, the configurations of the $1/2_1^+$ and $3/2_1^+$ states evolve to be predominantly characterized by $(1s_{1/2})^1(0d_{5/2})^4$. Compared to the ground state, the average occupancy of the $s_{1/2}$ orbital increases significantly in these states, resulting in larger MED values and more extended radial density distributions in the asymptotic region. Indeed, in the mirror nucleus $^{21}$Al, the proton $1s_{1/2}$ state is slightly unbound. Since $s$-waves lack a centrifugal barrier~\cite{LI2022137225,PhysRevC.84.051304,PhysRevC.101.031301}, the proton $1s_{1/2}$ wave function is more spatially extended compared to the deeply bound neutron $1s_{1/2}$ in $^{21}$O~\cite{PhysRevLett.113.142502,ensdf}, leading to a extend radial density distributions. Moreover, the stronger coupling involving the unbound $\pi 1s_{1/2}$ orbital in $^{21}$Al, which provides more binding energy than its neutron counterpart in $^{21}$O, results in a negative MED value. This means that the excitation energies of mirror states in proton-rich nuclei are lower than those in neutron-rich nuclei due to the strong couplings involving the unbound $s_{1/2}$ wave. Furthermore, our GSM calculations highlight the close relationship between MED and radial density distribution in mirror partners, offering insights for understanding the nature of isospin symmetry breaking.

\begin{figure}[htb!]
    \centering
    \includegraphics[width=1.\columnwidth]{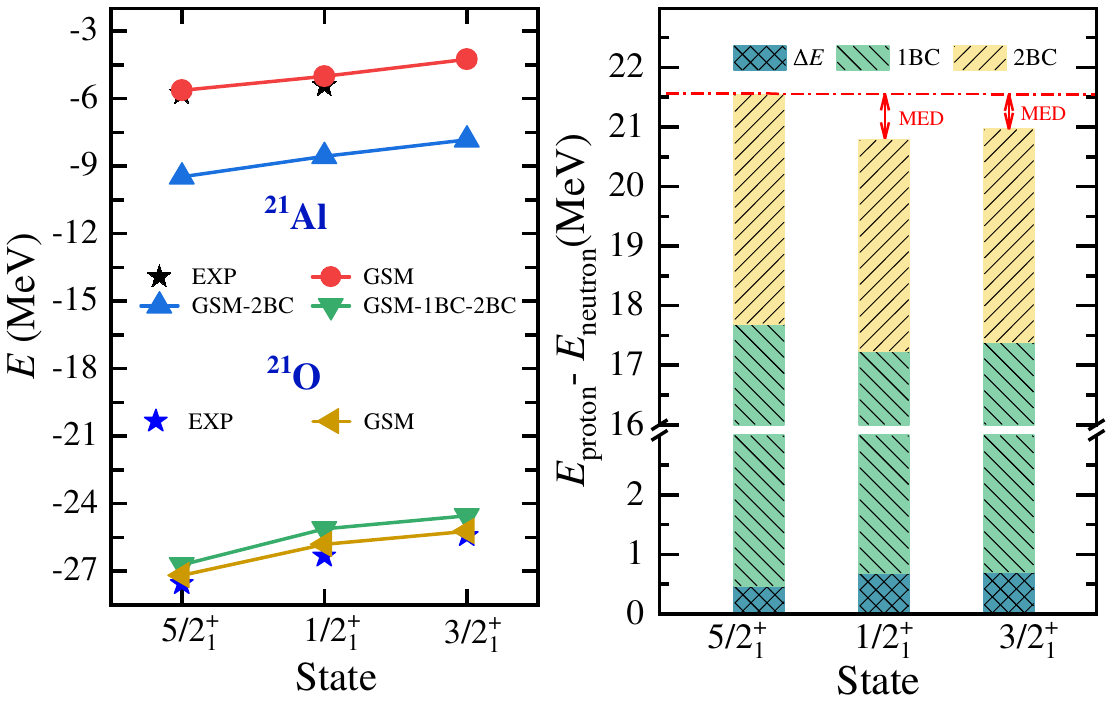}
    \caption{The calculated contribution of different components of the Hamiltonian for low-lying states in $^{21}$Al/$^{21}$O with GSM. The left panel shows the absolute energies relative to their inner core. 
GSM-2BC and GSM-1BC-2BC represent the calculated energies of $^{21}$Al minus two-body Coulomb force contribution and both one- and two-body Coulomb force contributions, respectively.
The right panel shows the value of each contribution, where $\Delta E$ is the difference between contribution of $NN$ interactions in $^{21}$Al/$^{21}$O. In the right panel, the red arrows indicate the energy differences between the ground and excited states of mirror nuclei, corresponding to the MED. Experimental data are taken from \cite{PhysRevC.110.L031301,ensdf}.}
    \label{21Al_21O_contribution}
\end{figure}

Additionally, we decomposed the GSM Hamiltonian into contributions from the nuclear interaction, the one-body Coulomb interaction between the inner core and valence protons (1BC), and the two-body Coulomb interaction between valence protons (2BC). We analyzed these components to investigate the origin of MED from the Hamiltonian perspective, as shown in Fig.~\ref{21Al_21O_contribution}. It is noteworthy that the GSM-1BC-2BC value represents the contribution from nuclear interactions. In the framework of isospin symmetry, the energy difference between mirror states should arise exclusively from Coulomb interactions, and the GSM-1BC-2BC values for a given state in a proton-rich nucleus should be close to the obtained energies of the neutron-rich mirror counterpart, as given in Ref.~\cite{arxiv}.
Our calculations indicate that the GSM-1BC-2BC value of the ground state in $^{21}$Al is very close to the GSM value in $^{21}$O with a derivation about 460 keV. However, for the ${1/2}^+$ and ${3/2}^+$ states, this difference becomes significant, which is about 680 and 695 keV respectively. To quantify this difference, we define $\Delta E$ as the difference between the GSM-1BC-2BC values in the proton-rich nucleus and the GSM values in its neutron-rich mirror partner, denoted as $\Delta E~=~\langle\Psi_{\mathrm{proton}}|H_{NN}|\Psi_{\mathrm{proton}}\rangle~-~\langle\Psi_{\mathrm{neutron}}|H_{NN}|\Psi_{\mathrm{neutron}}\rangle$, which $\Psi_{\mathrm{proton}}$ and $\Psi_{\mathrm{neutron}}$ are the many-body wave functions of rich-proton/neutron nuclei, as done in Ref.~\cite{arxiv}. The values of different parts to Hamiltonian are shown in Fig.~\ref{21Al_21O_contribution} right panel. Predominantly, the Coulomb interaction emerges as the dominant factor contributing to these $E(^{21}{\rm Al})-E(^{21}{\rm O})$ differences in all low-lying states,  and the $\Delta E$ making a minimal contribution.
But interestingly, the contributions of $\Delta E$, 1BC, and 2BC vary among different mirror states within each system were observed. The ${1/2}^+$ and ${3/2}^+$ states in $^{21}$Al/$^{21}$O show larger $\Delta E$ contributions and reduced Coulomb interactions compared to their ground states. The increased $\Delta E$ values highlight the significant role of nuclear interactions in isospin symmetry breaking, illustrating the complex competitive effects of forces that influence the energies of mirror nuclei.




    


    

\begin{figure}[htb!]
    \centering
    \includegraphics[width=0.88\columnwidth]{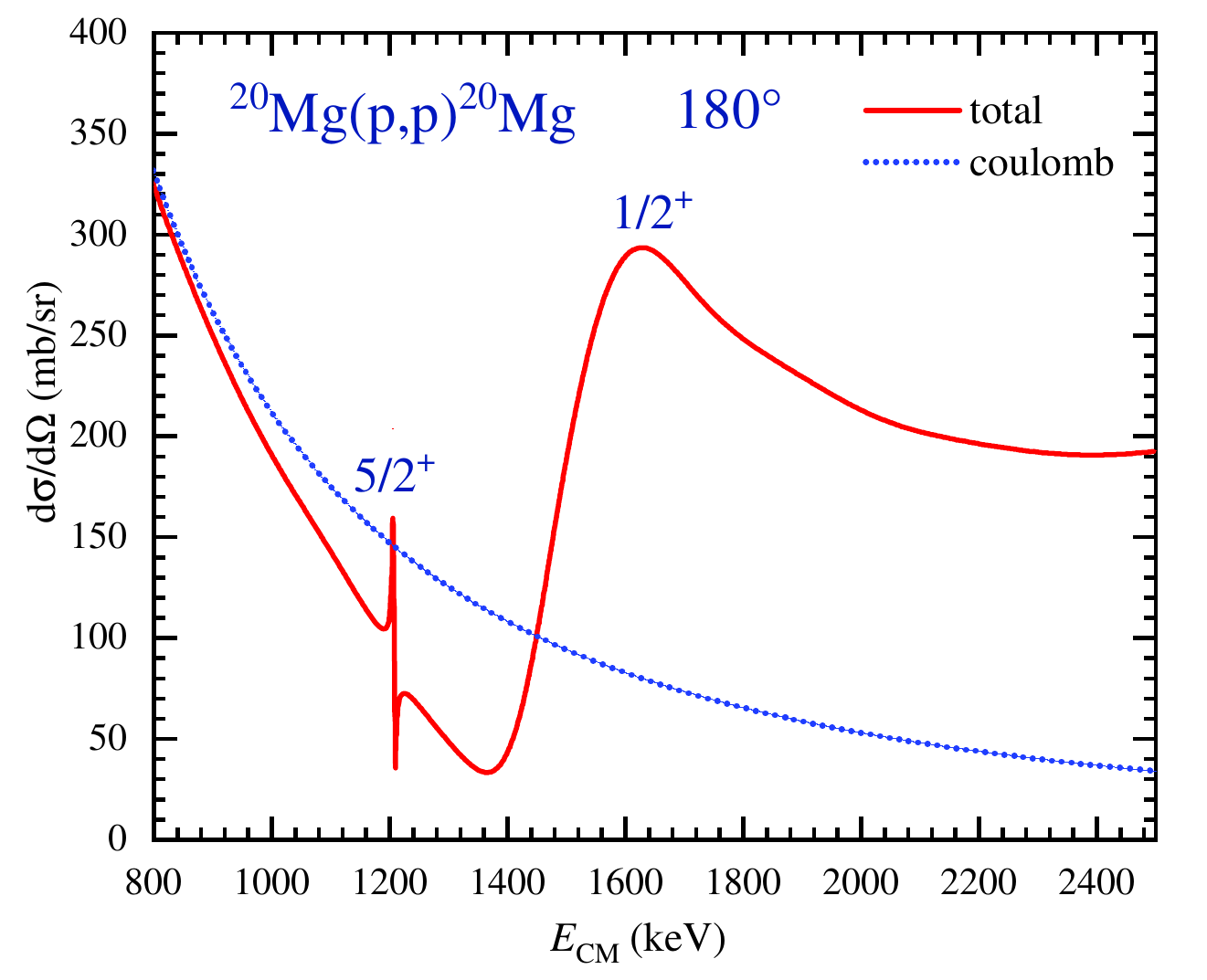}
    \caption{Excitation function of the $^{20}$Mg(p, p) scattering reaction calculated with GSM-CC. The energy and cross section are defined in the center of mass frame, with the cross section angle set at 180 degrees. The low-lying resonant states of $^{21}$Al are labeled next to their respective resonant peaks in the figure.}
    \label{20Mg(p, p)}
\end{figure}

The nuclear structure and reactions properties of  $^{21}$Al  are also ivestigated using GSM-CC approach. 
In the GSM-CC framework, the channel states of $^{21}$Al are constructed by coupling the ground state ($J^\pi = 0^+_1$) and the excited states ($J^\pi = 2^+_{1,2}$, $4^+_{1,2}$, and $0^+_2$) of $^{20}$Mg with the proton projectile in the following partial waves: $s_{1/2}$, $p_{1/2, 3/2}$, $d_{3/2, 5/2}$, and $f_{5/2, 7/2}$. 
The resulting composite states include the ground state $5/2^+$ and the first excited state $1/2^+$ of $^{21}$Al.
For the ground state $5/2^+$ of $^{21}$Al, the main coupling channels are $^{20}\text{Mg}(4^+) \otimes \pi d_{5/2}$, $^{20}\text{Mg}(2^+) \otimes \pi d_{5/2}$, and $^{20}\text{Mg}(0^+) \otimes \pi d_{5/2}$, contributing 52.5\%, 27.3\%, and 12.6\% to the total wave function, respectively. 
The most probable decay mode for this state is to the ground state of $^{20}\text{Mg}$ with the emission of a $d$-wave proton, resulting in a decay width of 0.385 keV. 
For the first excited state $1/2^+$, the dominant coupling channel is $^{20}\text{Mg}(0^+) \otimes \pi s_{1/2}$, contributing 72.8\%. This state is most likely to decay to the ground state of $^{20}\text{Mg}$, emitting an $s$-wave proton with a decay width of 225.257 keV. Furthermore, by performing GSM-CC calculations, we obtained the proton separation energy value $S_p=-1.19$ MeV, which is in good agreement with the experimentally determined value of $-1.15^{+0.10}_{-0.07}$ MeV~\cite{PhysRevC.110.L031301}. 
GSM-CC calculation also gives that the $1/2^+$ state has a relative energy of 1.49 MeV, resulting the excitation energy of the $1/2^+$ state is 0.3 MeV above the ground state, which matches perfectly with experimental data~\cite{PhysRevC.110.L031301}.




Following our calculations reproduce the level structure of $^{21}$Al, we then evaluated the excitation function for the $^{20}$Mg (p, p) reaction using the GSM-CC framework, providing a theoretical benchmark for future experiments. The calculated results are shown in Fig.~\ref{20Mg(p, p)}. The first peak, located at a resonance energy of approximately 1200 keV, corresponds to the $5/2^+$ narrow resonance state with a width of 0.38 keV. At a center-of-mass energy of around 1640 keV, the second peak aligns well with the broader $1/2^+$ state, which has a width of 225 keV. These predictions of the reaction cross sections are preliminary and require experimental validation in future studies.


\section {Summary} 

The study of unbound proton-rich nuclei beyond the proton drip line provides critical insights into nuclear stability and interactions. The recent discovery of a new isotope $^{21}$Al has emphasized the need for theoretical studies. Using the GSM and GSM-CC, we performed a comprehensive theoretical analysis of $^{21}$Al to elucidate its exotic features.
GSM calculations reproduce the ground state energies of the $A = 18-22$ isotones and their mirror partners. 
We also explored the excitation energies of $^{21}$Al/$^{21}$O, revealing large MEDs and significant TES, particularly in the $1/2^+$ states. 
To investigate isospin symmetry breaking in detail, we analyzed the many-body configurations, average occupations, radial density distributions, and the contribution of different components of the GSM Hamiltonian.
While isospin symmetry breaking is minimal in the ground states of $^{21}$Al/$^{21}$O, it becomes prominent in the excited states due to the increased occupation of the weakly bound or unbound $s_{1/2}$ orbital, leading to larger MEDs and extended radial density distributions. 
The differences in radial density distributions indicate variations in nuclear interactions, which are further highlighted by the differing nuclear interaction and Coulomb contributions across mirror states. 
These findings underscore the significant impact of TES and isospin symmetry breaking in mirror nuclei.

Furthermore, using the GSM-CC framework, we calculated the energies of the $5/2^+$ and $1/2^+$ states of $^{21}$Al, along with a proton separation energy, all of which show excellent agreement with experimental data. The $5/2^+$ ground state decays to the ground state of $^{20}$Mg via $d$-wave proton emission with a narrow decay width, while the $1/2^+$ excited state decays via $s$-wave proton emission with a broad decay width. The excitation function for the $^{20}$Mg(p, p) reaction is also predicted using GSM-CC, which gives two key resonances, corresponding to the $5/2^+$ and $1/2^+$ states. 
The results offer valuable insights into the exotic structure and reaction dynamics of $^{21}$Al, serving as a benchmark for future experimental studies.



\begin{acknowledgments}
We thank Nicolas Michel for the GSM code~\cite{MICHEL2020106978} used in the present work.
 This work has been supported by the National Key R\&D Program of China under Grant No. 2023YFA1606403; the National Natural Science Foundation of China under Grant Nos.  12205340, 12175281, 12347106, 12375123, 12475128,  and 12121005;  the Gansu Natural Science Foundation under Grant Nos. 22JR5RA123 and 23JRRA614; the Natural Science Foundation of Henan Province No. 242300421048; The numerical calculations in this paper have been done at Hefei Advanced Computing Center.
\end{acknowledgments}

\bibliographystyle{apsrev4-2} 
\bibliography{Ref}  

\end{document}